\documentclass[aps,prl,twocolumn,reprint,showpacs,superscriptaddress]{revtex4-1}
\usepackage{amsmath,amssymb,graphicx,color}

\newcommand\ket[1]{\left|\textstyle{#1}\right\rangle}
\newcommand\bra[1]{\left\langle\textstyle{#1}\right|}
\newcommand\braket[1]{\left\langle\textstyle{#1}\right\rangle}

\newcommand\adag{a^\dagger}

\begin{document}
\title{Universality in the Decay and Revival of Loschmidt Echoes}
\author{Myung-Joong Hwang}
\thanks{M.-J.H. and B.-B.W. contributed equally to this work.}

\affiliation{Institut f\"{u}r Theoretische Physik and IQST, Albert-Einstein-Allee 11, Universit\"{a}t Ulm, D-89069 Ulm, Germany}
\author{Bo-Bo Wei}
\thanks{M.-J.H. and B.-B.W. contributed equally to this work.}
\affiliation{School of Science and Engineering, The Chinese University of Hong Kong, Shenzhen, Shenzhen 518172, China}
\affiliation{Center for Quantum Computing, Peng Cheng Laboratory, Shenzhen 518055, China}
\author{Susana F. Huelga}
\affiliation{Institut f\"{u}r Theoretische Physik and IQST, Albert-Einstein-Allee 11, Universit\"{a}t Ulm, D-89069 Ulm, Germany}
\author{Martin B. Plenio}
\affiliation{Institut f\"{u}r Theoretische Physik and IQST, Albert-Einstein-Allee 11, Universit\"{a}t Ulm, D-89069 Ulm, Germany}

\begin{abstract}
A critically enhanced decay of the Loschmidt echo is characteristic of sudden quench dynamics near a quantum phase transition. Here, we demonstrate that the decay and revival of the Loschmidt echo follows power-law scaling in the system size and the distance from a critical point with equilibrium critical exponents. We reveal such dynamical scaling laws by analyzing relevant perturbations to the Loschmidt echo cast in a scaling invariant form. We confirm the validity and the generality of the predicted dynamical scaling laws with a diverse range of critical models such as Ising spin models with a short and long range interaction, a finite-component system phase transition, and a topological phase transition. Moreover, using the integrability of systems in the thermodynamic limit, we derive such scaling laws analytically from a microscopic analysis. Our finding promotes the Loschmidt echo to a quantitative non-equilibrium probe of criticality and allows for quantitative predictions on the role of criticality on various physical scenarios where the Loschmidt echo is central to describing non-equilibrium dynamics.
\end{abstract}
\maketitle\emph{Introduction.---}
Physical properties of systems undergoing phase transitions exhibit non-analytic behavior at a critical point~\cite{Sachdev:2011qu}. In addition to local order parameters and correlators of physical observables, through which phase transitions are typically characterised~\cite{Sachdev:2011qu}, the wave function itself exhibits an abrupt change near critical points of phase transitions, which can be captured by the quantum fidelity~\cite{Zanardi:2006fba}, an overlap between two ground states, and its susceptibility~\cite{You:2007ew}. Remarkably, the fidelity and fidelity susceptibility exhibit scaling laws like thermodynamic observables near critical points and such scaling laws make the quantum fidelity a general probe of phase transitions and their universality~\cite{Venuti:2007il,Zanardi:2007be,Yang:2007gr,Gu:2008fid,Tzeng:2008sc,Chen:2008in,Yang:2008fi,Kwok:2008qu,Abasto:2008fi,Zhao:2009si,Schwandt:2009jl,
Albuquerque:2010qu,Gu:2010fd,Rams:2011qu,Langari:2012qu,Mukherjee:2012quantum,Damski:2013fi,Sun:2015fi,Wang:2015bv,Sun:2017fid,Wei:2018fi}.

A Loschmidt echo (LE), defined as the fidelity between an initial ground-state and its time-evolved state under a quenched Hamiltonian~\cite{Jalabert:2001en}, is a dynamical analogue of the ground-state fidelity, whose behavior is also strongly influenced by the presence of a critical point~\cite{Quan:2006fda,Zhang:2008de,Zhang:2009di}. When both initial and final parameters of a quench protocol are near a critical point, the LE exhibits oscillatory behavior consisting of a rapid decay followed by a revival~\cite{Quan:2006fda,Yuan:2007z,Rossini:2007de,Zhang:2008de,Zhang:2009di,Zhong:2011ec,Happola:2012j,Montes:2012ph,Sharma:2012st1,Sharma:2012st2,Rajak:2014ra,Jafari:2017ib}.  The essential feature found is that the initial decay becomes progressively sharper and the revival period longer as one moves closer to the critical point or increase the system size, for a fixed quench size. While this so-called critically enhanced decay of LE has been suggested as a probe of phase transitions~\cite{Quan:2006fda,Zhang:2008de,Zhang:2009di}, its wide applicability has been hindered by a lack of quantitative scaling laws, analogous to those derived earlier for the quantum fidelity. Moreover, as the LE is a central notion to characterize a variety of phenomena in non-equilibrium dynamics ranging from decoherence in the central spin model~\cite{Cucchietti:2003de,Yang:2016qm}, the effects of non-Markovianity~\cite{Haikka:2012jp}, and the statistics of quantum work distributions~\cite{Dorner:2013ex,Mazzola:2013me}, such quantitative scaling laws for the LE could shed new light on the role of criticality in these non-equilibrium dynamics.

In this letter, we establish universal dynamical scaling laws for the critical decay of LE and thereby promote the LE as a \emph{quantitative} non-equilibrium probe of criticality. This is achieved in two steps, connecting two seemingly disparate properties of LE near a phase transition: (i) We first show that a scaling transformation based on scaling dimensions of relevant parameters leave the equation of motion for LE and thus the LE itself invariant. (ii) We then identify the critical decay of LE as a consequence of perturbatively breaking its scaling invariance due to either reduced system size or increased quench size away from the critical point. Using this strategy, we find dynamical scaling functions of LE and derive a power-law decay of LE for increasing system sizes and decreasing distances from a critical point, solely in terms of equilibrium critical exponents.

 We then use various critical models to confirm the validity and the generality of the predicted scaling laws for the LE. The considered models range from the transverse field Ising chain (TFIC)~\cite{Suzuki:2012qu} and the Lipkin-Meshkov-Glick (LMG) model~\cite{Lipkin:1965va}, which are spin models with a short and long range interaction, respectively, to the quantum Rabi model (QRM)~\cite{Rabi:1936pr} which is a paradigmatic example of a finite-component system phase transition of coupled spins and bosons~\cite{Hwang:2015eq,Hwang:2016cb,Puebla:2017gq}. In particular, using the LMG model and the QRM, we derive analytically the scaling laws from the microscopic models that agree with the general scaling analysis and also show that the LE is lower-bounded by the ground-state fidelity. In addition, we consider the Su-Schrieffer-Heeger (SSH) model~\cite{Su:1979so} to demonstrate the existence of scaling laws of LE for a topological phase transition.

\emph{Scaling laws of the Loschmidt echo.---}
We consider a family of Hamiltonians $H(\lambda)=H_0+\lambda H_1$ controlled by $\lambda$, and a sudden quench protocol where $\lambda$ is changed from $\lambda_i$ to $\lambda_f$ and the initial state $\ket{\psi_0(\lambda_i)}$ is the ground state of $H(\lambda_i)$. The LE of such a quench protocol is defined as
\begin{equation}
\label{eq01}
	L(N,\lambda_i,\lambda_f,t)=|\braket{\psi_0(\lambda_i)|e^{-itH(\lambda_f)}|\psi_0(\lambda_i)}|^2.
\end{equation}
%Note that the system size $N$ is explicitly indicated only for the LE and is implicitly assumed for other quantities.

Suppose that $H(\lambda)$ undergoes a second-order quantum phase transition (QPT) at $\lambda=\lambda_c\equiv1$ in the thermodynamic limit ($N\rightarrow\infty$), characterized by a diverging correlation length $\xi\sim |\lambda-\lambda_c|^{-\nu}$ and a vanishing energy gap $\Delta\sim |\lambda-\lambda_c|^{\nu z}$ with $\nu$ and $z$ being the critical exponents. We define $\lambda_i=\lambda_c-\delta \lambda$ and $\lambda_f=\lambda_i-g$ and assume $\delta\lambda,g>0$ for a simplicity. The goal of this section is to establish scaling laws for the critical decay of LE, solely determined by the critical exponents $z$ and $\nu$. Below, we use standard scaling arguments and techniques to arrive at proposed scaling relations. While these techniques may be made rigorous by a renormalisation group analysis, we verify the correctness of the so obtained scaling relations for the LE at the hand of concrete examples in the next sections.

The first step is to establish a scaling invariance of the LE near the critical point through its equation of motion~\cite{Nikoghosyan:2016ff}, which reads
 \begin{equation}
\label{eq02}
\frac{dL(N,\delta \lambda, g,t)}{dt}=-i\braket{\psi_0(\lambda_i)| [H(\lambda_f),\rho(t)]|\psi_0(\lambda_i)},
\end{equation}
where $\rho(t)=e^{-itH(\lambda_f)}|\psi_0(\lambda_i)\rangle\langle\psi_0(\lambda_i)|e^{itH(\lambda_f)}$ and $L(t=0)=1$. In a scaling limit of $N^{-1},\delta\lambda, g\ll1$, the most remarkable aspect of a critical system is that a renormalization of the system size $N\rightarrow N/b$ leaves the partition function invariant~\cite{Cardy:1996th}. This fact dictates how system parameters and operators ($O$) should scale upon the renormalization of $N\rightarrow N/b$, which are expressed in terms of the scaling dimensions $[O]$, defined as  $O\rightarrow O b^{[O]}$. Using the known scaling dimensions~\cite{Sachdev:2011qu,Cardy:1996th}, let us define a scaling transformation,
\begin{align}
\label{eq03}
\tilde{N}=b^{-1}N,\quad \tilde{t}=b^{-z}t,\quad  \tilde {\delta\lambda} =b^{1/\nu}\delta\lambda ,\quad  \tilde g =b^{1/\nu} g.
\end{align}
In addition, we have the scaling dimension of the critical Hamiltonian, $\tilde{H}(\lambda_c)=b^{z}H(\lambda_c)$ and $\tilde{H_1}=b^{z-1/\nu}H_1$~\cite{Schwandt:2009jl}. %We note that the following analysis could be generalized to cases where $g$ couples to a different perturbative Hamiltonian than $H_1$ given the knowledge of its scaling dimension.  
Under the transformation Eq.~\eqref{eq03}, the equation of motion Eq.~\eqref{eq02} and the initial condition $L(t=0)=1$ are invariant and therefore the LE has a scaling invariance,
\begin{align}
\label{eq04}
L(N,\delta\lambda,g,t)=L(Nb^{-1},\delta\lambda b^{1/\nu},g b^{1/\nu},t b^{-z}).
\end{align}
The above scaling invariance agrees with an earlier conjecture and numerical results~\cite{Quan:2006fda,Haikka:2012jp,Pelissetto:2018ic}. Below, we analytically derive the above scaling invariance from a microscopic analysis of a certain class of models, thereby confirming its validity for all parameter ranges beyond the reach of numerical calculations.

The second step is to connect the scaling invariance of the LE to its critically enhanced decay, which at first sight don't seem to be related as the minimum of the LE remains unchanged under the transformation, thus no critically enhanced decay. Suppose, however, that we repeat the renormalization procedure $n$ times, $L(N,\delta\lambda,g,t)=L(Nb^{-n},\delta\lambda b^{n/\nu},g b^{n/\nu},t b^{-nz})$. This system remains in the scaling limit only when the rescaled parameters continue to satisfy $\frac{1}{N}b^{n},\delta\lambda b^{n/\nu},g b^{n/\nu}\ll1$. The first parameters to exceed unity defines the most relevant perturbation breaking of the scaling invariance~\cite{Cardy:1996th,Hauke:2016ht} and we show that this leads to the critical decay of LE.

We first examine $\delta\lambda,g \ll 1/N$, in which case $1/N$ becomes relevant perturbation when $Nb^{-n}\sim1$. Therefore we substitute $b^{n}$ by $N$ in Eq.~\eqref{eq04}, which leads to a dynamical scaling function
\begin{eqnarray}
\label{fssregime}
L(N,\delta\lambda,g,t)&=&\Psi_1\left(N^{1/\nu}\delta\lambda,N^{1/\nu}g, N^{-z}t\right).
\end{eqnarray}
The minimum of the LE satisfies $\displaystyle L_\textrm{min}(N,\delta\lambda,g)=\min_{D}\Psi_1\left(N^{1/\nu} \delta\lambda,N^{1/\nu}g,D\right)=\Psi_1^{\text{min}}\left(N^{1/\nu} \delta\lambda,N^{1/\nu}g\right)$ with $D\equiv N^{-z}t=D_\textrm{min}$. From $D_\textrm{min}=N_1^{-z}t_\textrm{min,1}=N_2^{-z}t_\textrm{min,2}$, $t_\textrm{min}$ for different $N$ are related by $t_\textrm{min,1}=t_\textrm{min,2}(N_1/N_2)^{-z}$. For $\delta\lambda=0$, using the fact that $\Psi_1^\textrm{min}=1$ at $g=0$ (no quench) and the Taylor expansion of  $\Psi_1^\textrm{min}$ for $N^{1/\nu}g\ll1$, we find 
\begin{eqnarray}
\label{eq05}
1-L_\textrm{min}(N,g)=1-\Psi^\textrm{min}_1\left(0,N^{1/\nu}g\right)\propto (gN^{1/\nu})^2.
\end{eqnarray}
This power law governs the onset of the critical enhancement of the LE decay as one increases $N$.

\begin{figure}[t]
\centering
\includegraphics[width=\linewidth,angle=0]{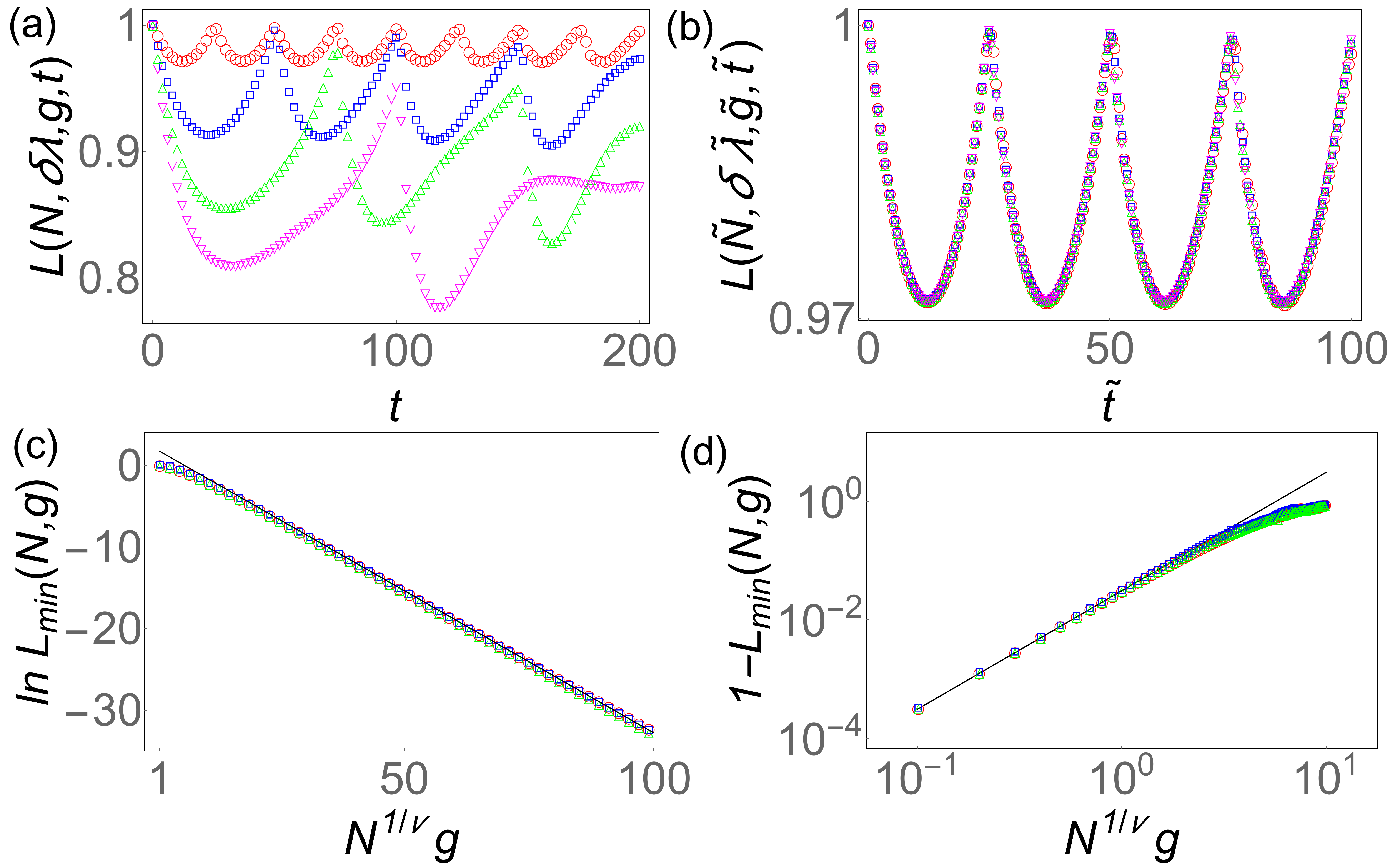}
\caption{Universal scaling of the Loschmidt echo (LE) in the transverse field Ising chain. (a). The LE $L(N,\delta\lambda,g,t)$ with $\delta\lambda=0.02$ and $g=0.01$ as a function of time for different system sizes, $N=100$, $200$, $300$ and $400$ (from top to bottom). (b) Scaling invariance of LE given in Eq.~\eqref{eq04}. We take $N=100,\delta\lambda=0.02,g=0.01$ and $\nu=z=1$. All the data for $b=1$, $1/2$, $1/3$, and $1/4$ collapses onto a single curve. (c,d) The data collapse of the LE minimum as a function of $N^{1/\nu}g$ with $\delta\lambda=0$ for different values of $N$ and $g$ satisfying (c) $N^{1/\nu}g\gg1$ and (d) $N^{1/\nu}g\ll1$.}
\label{fig:1}.
\end{figure}

Next, we consider $\frac{1}{N},\delta\lambda^{\nu}\ll g^{\nu}$ where $g$ represents the most relevant perturbation for $gb^{n/\nu}\sim1$. Therefore, by trading $b^{n}$ for $g^{-\nu}$ in Eq.~\eqref{eq04}, we find another dynamical scaling function,
\begin{eqnarray}\label{gdominate}
L(N,\delta\lambda,g,t)&=&\Psi_2\left(N^{-1}g^{-\nu},g^{-1}\delta\lambda,g^{\nu z}t\right),
\end{eqnarray}
from which we define $\displaystyle L_\textrm{min}(N,\delta\lambda,g)=\min_{\bar D}\Psi_2\left(N^{-1}g^{-\nu},g^{-1}\delta\lambda,\bar D\right)=\Psi_2^{\text{min}}\left(N^{-1}g^{-\nu},g^{-1}\delta\lambda\right)$. In this case, we have $t_\textrm{min,1}=t_\textrm{min,2}(g_2/g_1)^{z\nu}$. At $\delta\lambda=0$, the scaling variable of $\Psi^\textrm{min}_1$ and $\Psi^\textrm{min}_2$ are identical, i.e., $N^{1/\nu}g$, and therefore we have a single scaling function for both parameter regime, $\Psi^\textrm{min}\equiv\Psi^\textrm{min}_1|_{\delta\lambda=0}=\Psi^\textrm{min}_2|_{\delta\lambda=0}$. In the next section, we confirm the existence of the single dynamical scaling function $\Psi^\textrm{min}(N ^{1/\nu}g)$ by numerical results. Unlike the model independent asymptotic scaling in Eq.~\eqref{eq05}, we find that critical models may exhibit qualitatively different asymptotic scaling of the $\Psi^\textrm{min}(N ^{1/\nu}g)$ for $N^{1/\nu}g\gg1$. For the critical models with a few effective degrees of freedom, such as LMG model and QRM, in the next section, we find the finite-size scaling for the critically enhanced decay of the LE for $N\gg1$,
\begin{eqnarray}
\label{eq06}
L_\textrm{min}=\Psi^\textrm{min}(N^{1/\nu}g)\propto N^{-z}\quad\textrm{for}\quad N^{1/\nu}g\gg1,
\end{eqnarray}
and the critical scaling in the thermodynamic limit,
\begin{eqnarray}
\label{eq07}
\lim_{N\rightarrow\infty}L_\textrm{min}\propto \delta \lambda^{z\nu}.
\end{eqnarray}
On the other hand, for the interacting models such as TFIM and SSH model where there are many modes participating in the LE dynamics, we find an exponential decay of LE [Fig.~\ref{fig:1}-(c)].

\emph{Implications---}  Let us briefly remark on the important implications of our finding before confirming its validity using concrete examples. We note that the LE can be experimentally measured for a generic critical system by probing the decoherence dynamics of a single qubit that is globally coupled to the critical model~\cite{Quan:2006fda,Zhang:2008de,Zhang:2009di,Yang:2016qm}. Therefore, the scaling laws of LE, Eq.~\eqref{eq04}-\eqref{eq07}, allow one to extract universal properties of QPTs, such as the location of QCP, all the critical exponents and the universal scaling functions, from measured LE data without any prior knowledge on the nature of critical point. While we considered the $T=0$ case, the scaling laws survive for a finite-temperature $T$ by a proper rescaling of temperature $T\rightarrow b^zT$ at a low temperature and by a dynamical decoupling at a high temperature~\cite{Wei}. Moreover, the LE is directly linked to the quantum Fisher information~\cite{Fiderer:2018hc}, which underpins recently proposed measures of non-Markovianity and can be related to information backflow~\cite{Lu:2010bn,Breuer:2016dn,Rivas:2014bl}. On the basis of our result, the quantitative relation for non-Markovianity and criticality observed in Ref.~\cite{Haikka:2012jp} can now be argued to hold for any critical environment. Finally, as the LE is the characteristic function of the quantum work distribution in a sudden quench process~\cite{Dorner:2013ex,Mazzola:2013me}, the scaling laws for LE also implies that the quantum work distribution for a critical system is universal.

\begin{figure}[t]
\centering
\includegraphics[width=0.46\linewidth,angle=0]{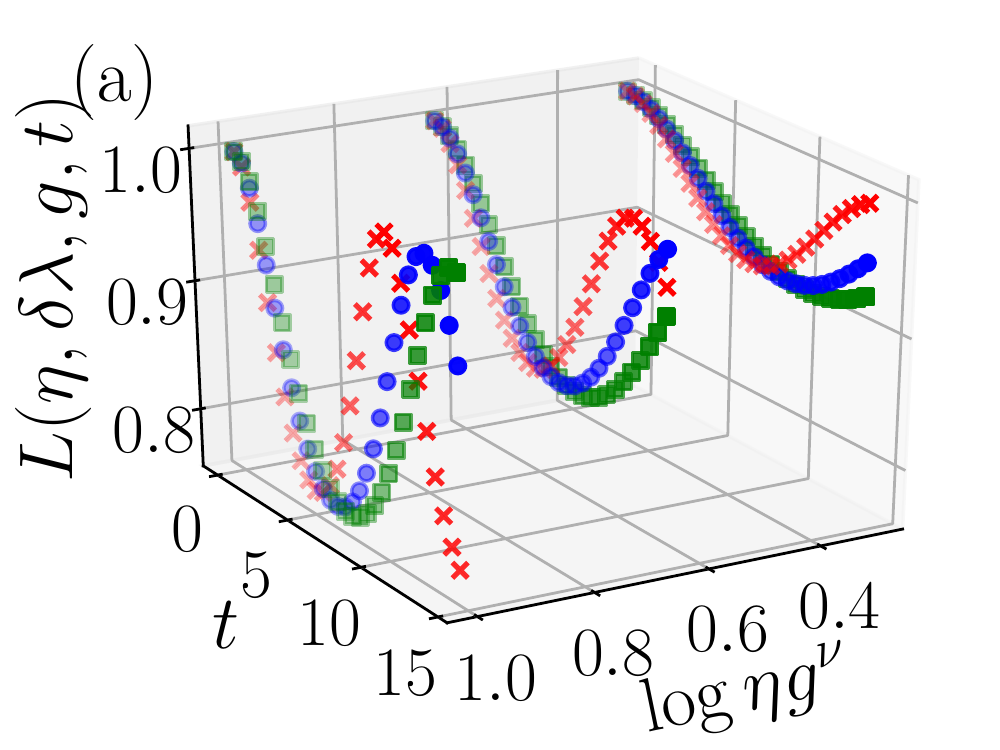}
\includegraphics[width=0.46\linewidth,angle=0]{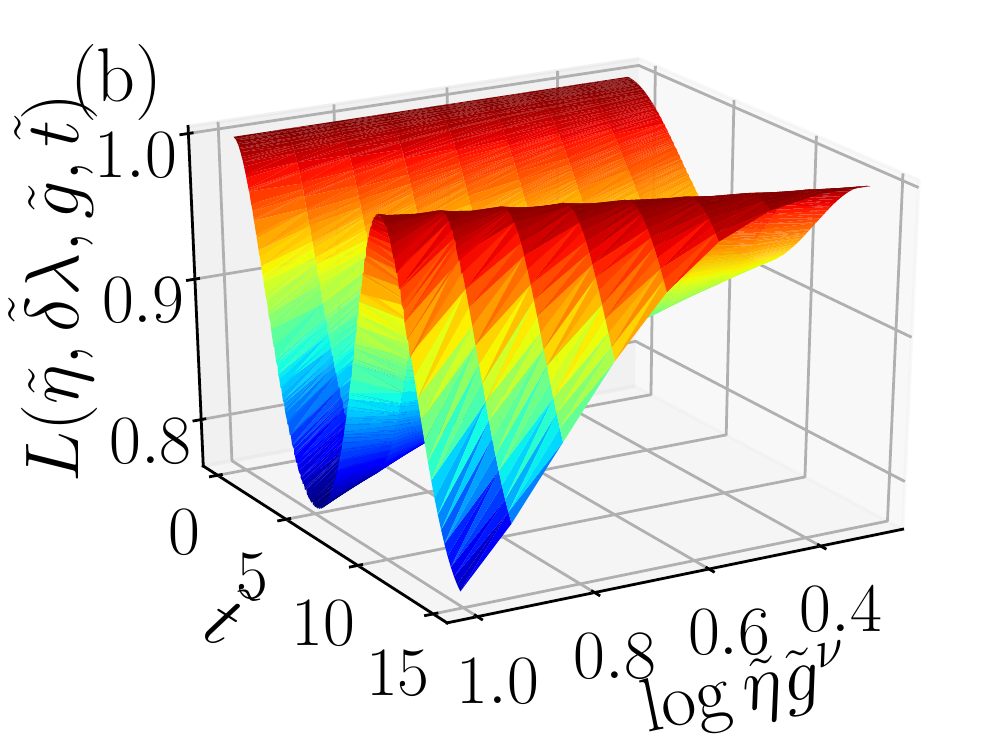}\\
\includegraphics[width=0.46\linewidth,angle=0]{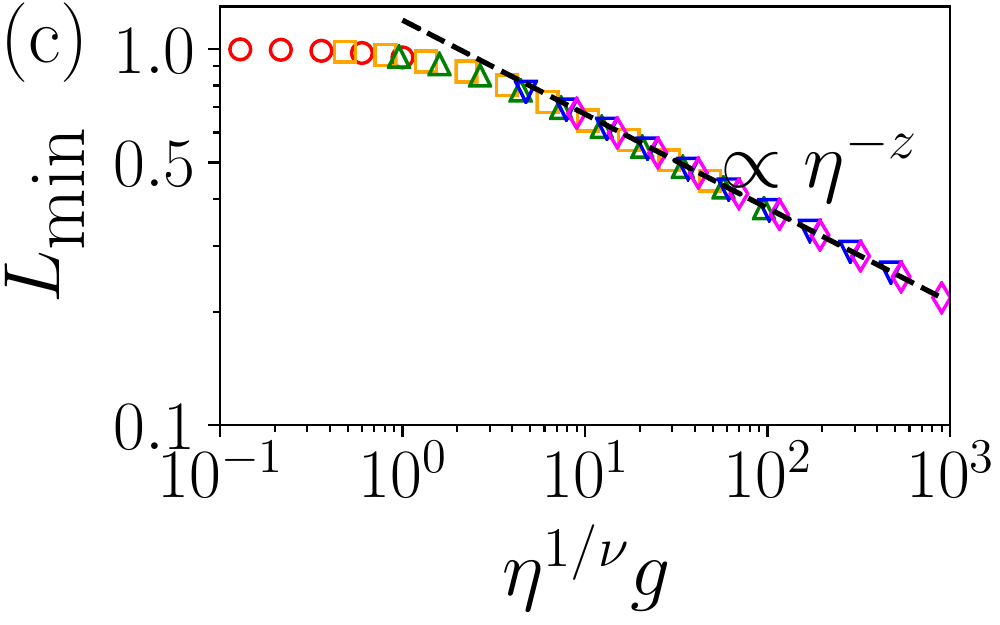}
\includegraphics[width=0.46\linewidth,angle=0]{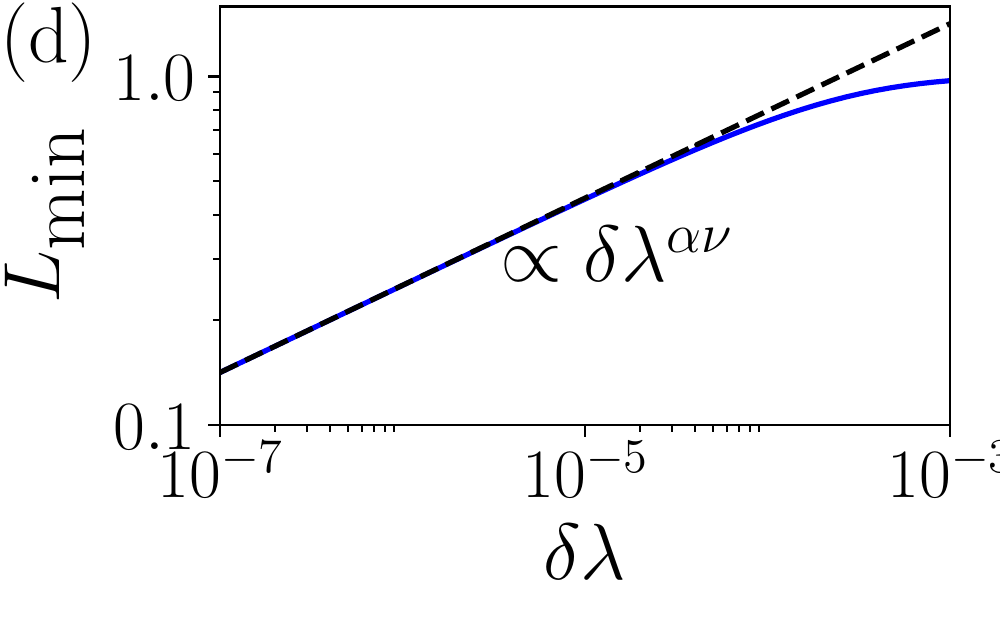}
\caption{Universal scaling of the LE in the QRM. (a) The LE at $\delta\lambda=0$ for $\eta g^\nu=2$, $4.47$, $10$. For each $\eta^{1/\nu}g$, the LE for $\eta=1000/b$ with $b=1$, $1/2$, and $1/3$ are presented (b) Dynamical scaling function of LE. For a rescaled time $\tilde t =t b^{-z}$, all data for different $\eta$ and $g$ collapse onto a single surface. (c) Data collapse of $L_\textrm{min}$ at $\delta\lambda=0$ as a function of $N^{1/\nu}g$. The power-law decay at $\eta g^\nu\gg1$ agrees with the predicted $\eta^{-z}$. (d). The critical scaling of $L_\textrm{min}$ in the limit $\eta\rightarrow\infty$ with an exponent $\nu z=1/2$.}
\label{fig:2}.
\end{figure}

\emph{Transverse field Ising models.---}
Now we turn our attention to concrete critical models to show the applicability of the predicted scaling laws. First we consider Ising models in a transverse field \cite{Suzuki:2012qu}, namely,
\begin{eqnarray}
H=-\sum_{i,j=1}^N[J_{ij}\sigma_i^x\sigma_j^x +J\lambda\sigma_j^z].
\end{eqnarray}
Here $\sigma_j^{x,y,z}$ denotes the Pauli operator for $j$th spin, $J_{ij}$ the interaction between $i$th and $j$th spins, $\lambda$ the strength of transverse field. We will consider two limiting cases of both short and long-range interactions; the first is the transverse field Ising chain (TFIC) with $J_{ij}=J\delta_{i+1,j}$ and the second is the Lipkin-Meshkov-Glick (LMG) model with $J_{ij}=J/N$. Both models exhibit a QPT at $\lambda=1$ that belong to different universality classes. Namely, TFIC belongs to the Ising universality class with critical exponents $\nu=z=1$~\cite{Sachdev:2011qu} and the LMG model to the class of infinitely-coordinated models with $\nu=3/2$ and $z=1/3$~\cite{Botet:1982ju}, which includes QRM considered below.

In Fig.~\ref{fig:1}, we present numerical solutions for the LE dynamics of TFIC. A typical LE decay and revival dynamics are shown in Fig.~\ref{fig:1}-(a) where the decay of the LE is enhanced as $N$ increases. For a set of parameters related by the scaling transformation in Eq.~\eqref{eq03}, the LE dynamics collapses onto a single curve [Fig.~\ref{fig:1}-(b)], showing the scaling invariance in Eq.~\eqref{eq04}. Finally, in Fig.~\ref{fig:1}-(c,d), we present $L_{\text{min}}$ for different values of $N$ and $g$ at $\delta\lambda=0$ as a function of the rescaled parameter $N^{1/\nu}g$. It shows a perfect data collapse which confirms the existence of the dynamical scaling function $\Psi_\textrm{min}(N^{1/\nu}g)$ predicted in the previous section. Moreover, for $N^{1/\nu}g\gg1$, we observe an exponential decay [Fig.~\ref{fig:1}-(c)] with a numerical fit $\ln L_{\text{min}}=2.10-0.35\times(N^{1/\nu}g)$, which can be attributed to the fact that the number of modes participating in LE dynamics of TFIC is extensive. Finally, the asymptotic behavior of $1-L_\textrm{min}\propto (N^{1/\nu}g)^{2/\nu}$, as predicted in Eq.~\eqref{eq05}, is confirmed [Fig.~\ref{fig:1}-(d)]. In addition to TFIC, we also confirm that all the scaling laws hold for the long-range interaction (LMG model) and we refer readers to Supplementary Materials~\cite{sup} for the analysis of LE dynamics including the derivation of scaling laws including power-law scaling in the limit $N^{1/\nu}g\gg1$ given in Eqs.~\eqref{eq06} and \eqref{eq07}, as well as a proof that the LE is lower-bounded by the ground-state fidelity.

\begin{figure}[t]
\centering
\includegraphics[width=\linewidth,angle=0]{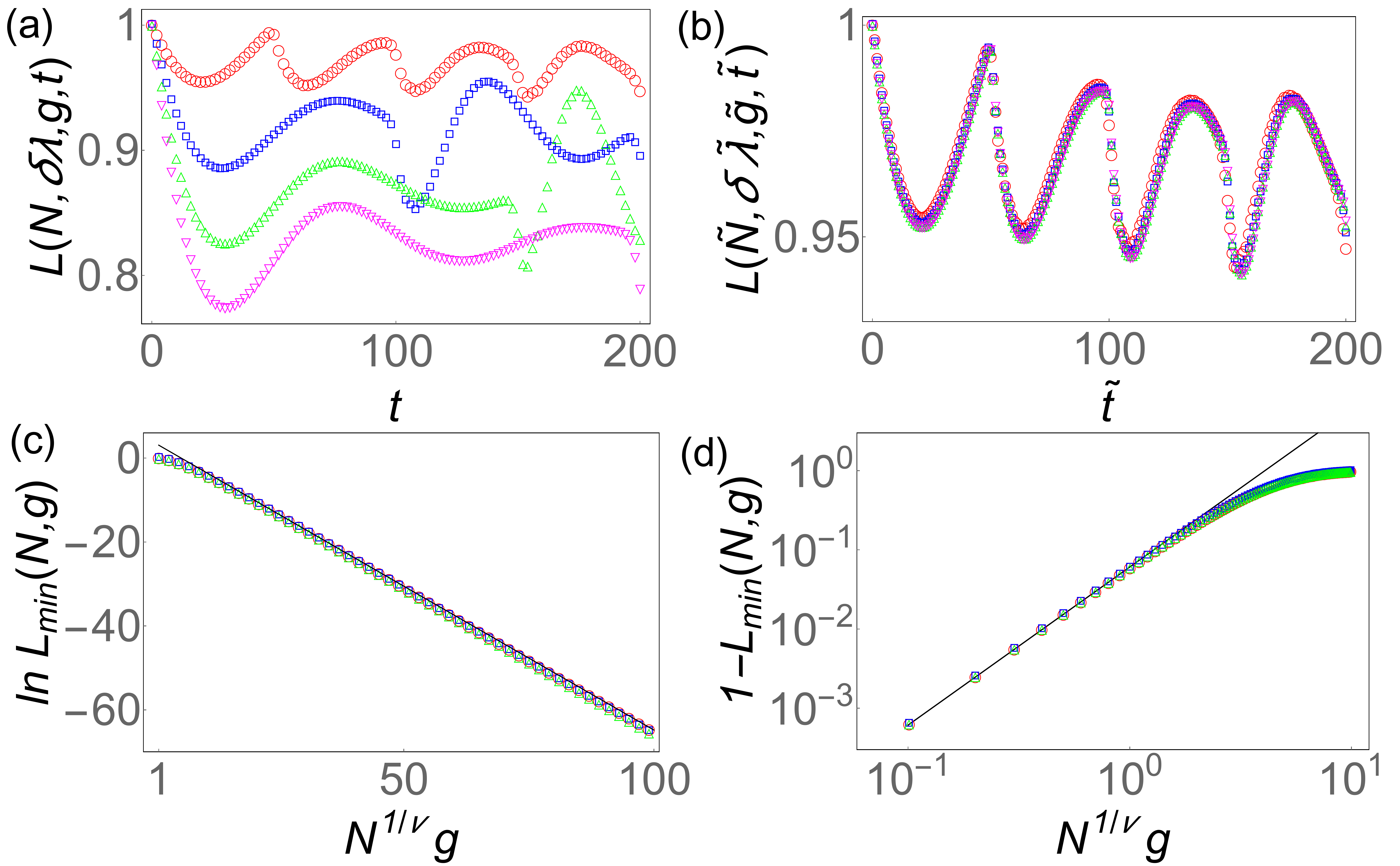}
\caption{Universal scaling of the LE in the SSH model (a) The LE dynamics as a function of time for different system sizes (b) Scaling invariance of LE. (c,d) The data collase of the LE minimum as a function of $N^{1/\nu}g$ with $\delta\lambda=0$ for different values of $N$ and $g$ satisfying (c) $N^{1/\nu}g\gg1$ and (d) $N^{1/\nu}g\ll1$. All parameters used here are identical with the one for Fig.~\ref{fig:1} and the Ising critical exponent $z=\nu=1$ are used for scaling transformations.}
\label{fig:3}.
\end{figure}

\emph{The quantum Rabi model.---} Here we consider the QRM~\cite{Rabi:1936pr},
\begin{eqnarray}
\label{QRM}
H_\textrm{QRM}=a^\dagger a +\frac{\eta}{2}\sigma_z+\frac{\lambda\sqrt{\eta}}{2}\sigma_x(a+a^\dagger)
\end{eqnarray}
where $\sigma_{x,z}$ are Pauli operators and $a$ a boson operator. $\eta$ and $\lambda$ are the transition frequency and the coupling strength, respectively, in unit of the boson frequency. The QRM undergoes a QPT at $\lambda=1$ in the limit of $\eta\rightarrow\infty$~\cite{Hwang:2015eq}. It belongs to the same universality class with the LMG model and the Dicke model and the role of system size $N$ is played by $\eta$, which can be controlled in experiments to reach the scaling limit~\cite{Puebla:2017gq}.

For $\eta\rightarrow\infty$, the effective Hamiltonian of the QRM becomes quadratic~\cite{Hwang:2015eq} and we find the exact analytical expression for the LE as
\begin{align}
\label{qrm01}
L(\delta\lambda,g,t)=|\cosh^2(z_0)-e^{2it\epsilon(1+\delta\lambda+g)}\sinh^2(z_0)|^{-1},
\end{align}
where $z_0=\frac{1}{4}\ln[\left(1-(1+\delta\lambda+g)^2\right)/\left(1-(1+\delta\lambda)^2\right)]$ and $\epsilon(\lambda)=\sqrt{1-\lambda^2}$. The first minimum of Eq.~\eqref{qrm01} is $L_\textrm{min}\propto|\delta\lambda+g|^{z\nu}$,
which occurs at $t_\textrm{min}=\pi/\epsilon(1+\delta\lambda+g)\propto|\delta\lambda+g|^{-z\nu}$. Therefore, $L_\textrm{min}$ and $t_\textrm{min}$ vanishes and diverges, respectively, as $\delta\lambda,g\rightarrow0$. In order to restore the analyticity of $L_\textrm{min}$ for $\eta<\infty$, we express it as $L_\textrm{min}|_{g\rightarrow0}=|\delta\lambda|^{z\nu}G_\textrm{min}\left(\eta\delta \lambda^{\nu}\right)$ with $\lim_{x\rightarrow0}G_\textrm{min}(x)=x^{-z}$. This leads to the finite-size scaling $L_\textrm{min}\propto\eta^{-z}$ alluded in Eq.~\eqref{eq06}. Finally, in the supplementary material~\cite{sup}, we derive an expression for the LE for finite $\eta$ and show that they are invariant under the transformation Eq.~\eqref{eq03}, which constitutes an analytical confirmation of Eq.~\eqref{eq04}.

In Fig.~\ref{fig:2}-(a) and (b), we show the numerically calculated LE dynamics at $\delta\lambda=0$ as a function of $t$ and $\eta^{1/\nu}g$. For a fixed $\eta^{1/\nu}g$, increasing $\eta$ makes $t_\textrm{min}$ longer while keeping $L_\textrm{min}$ same. By rescaling of time according to Eq.~\eqref{eq04}, we find that the LE collapses on to a 2D surface, which corresponds to the dynamical scaling function $\Psi_1\left(\eta^{1/\nu}\delta\lambda=0,\eta^{1/\nu}g,\eta^{-z}t\right)$ in Eq.~\eqref{fssregime}. Furthermore, the scaling function $\Psi^\textrm{min}(\eta^{1/\nu}g)$ exhibits a power-law decay $\eta^{-z}$ with $z=1/3$ in the limit of $\eta^{1/\nu}g\gg1$ [Fig.~\ref{fig:2}-(c)] and we also observe $1-\Psi^\textrm{min}(\eta^{1/\nu} g)\propto\eta^{\nu/2}$ (not shown).  In the limit of $\eta\rightarrow\infty$, we find the vanishing of $L_{\text{min}}$ at $\delta\lambda$ is governed by a power-law $\delta\lambda^{z\nu}$ with $z\nu=1/2$ [Fig.~\ref{fig:2}-(d)]. Finally, we find that the LE is lower-bounded by the ground-state fidelity, i.e. $\displaystyle \min_t L(\lambda_i,\lambda_i-g,t)=(2 F(\lambda_i,\lambda_i+g)^{-4}-1)^{-1}$ where $F(\lambda_i,\lambda_i+g)\equiv \braket{\psi_0(\lambda_i)|\psi_0(\lambda_i-g)}$~\cite{sup}.

\emph{Su-Schrieffer-Heeger model.---} We examine whether the scaling law of the LE also holds for a topological phase transition (TPT), as exemplified by the Su-Schrieffer-Heeger (SSH) model~\cite{Su:1979so},
\begin{eqnarray}\label{ham}
H&=&\sum_{j=1}^N\left[J_1a_j^{\dagger}b_j+J_2b_j^{\dagger}a_{j+1}+H.c.\right].
\end{eqnarray}
Here $a_j$ and $b_j$ are the annihilation operators in two sublattices at site $j$, $J_{1(2)}$
are hopping amplitudes. It undergoes a TPT at $\lambda\equiv J_2/J_1=1$. We observe in Fig.~\ref{fig:3} that the LE of SSH model satisfies all the scaling laws when the Ising critical exponents $\nu=z=1$ are used. This can be understood from that the SSH model can be mapped to the quantum compass model~\cite{You:2014fp,Nussinov:2015co} whose QPT belongs to the Ising universality class. We note that the scaling behaviors are identical for quenches starting in topologically trivial or non-trivial phase. Therefore, we conclude that the LE can be used to probe the universal properties of TPTs, while it is insensitive to the topological order.

\emph{Conclusions---} We have established general scaling laws for the LE dynamics following a sudden quench and have confirmed its validity and generality using several paradigmatic critical models representing different types of phase transitions. The quantitative scaling laws allow one to probe criticality from the experimentally measurable LE dynamics~\cite{Quan:2006fda,Zhang:2008de,Zhang:2009di,Yang:2016qm} for a generic critical system without a prior knowledge on the nature of a critical point. Furthermore, the scaling laws of LE directly implies the universality of a wide range of non-equilibrium phenomena characterized by the LE, including the statistics of quantum work distribution and the quantitative relation between Markovianity and criticality~\cite{Haikka:2012jp}.

%\emph{Acknowledgement---} 
\begin{acknowledgements}
M.J.H., S.F.H., M.B.P were supported by the ERC Synergy grant BioQ. M.J.H. thanks Mark Mitchison for an insightful discussion. B.~B.~W.~was supported by the National Natural Science Foundation of China (Grant Number 11604220) and the President's Fund of The Chinese University of Hong Kong, Shenzhen. A part of numerical calculation is performed using QuTip~\cite{Johansson:2013gb}.
\end{acknowledgements}

\pagebreak
\widetext
\begin{center}
\textbf{ \large Supplemental Material: \\Supplementary Materials: Universality in the Decay and Revival of Loschmidt Echoe}
\end{center}
\setcounter{equation}{0}
\setcounter{figure}{0}
\setcounter{table}{0}

%\makeatletter
\renewcommand{\theequation}{S\arabic{equation}}
\renewcommand{\thefigure}{S\arabic{figure}}
\renewcommand{\bibnumfmt}[1]{[S#1]}
\renewcommand{\citenumfont}[1]{S#1}

\section{Section A: Loschmidt echo in the Lipkin-Meshkov-Glick model.}
The Hamiltonian of the Lipkin-Meshkov-Glick (LMG) model is,
\begin{eqnarray}
H(\lambda)&=&-\frac{J}{N}\sum_{1\leq i<j\leq N}\Big(\sigma_i^x\sigma_j^x+\gamma\sigma_i^y\sigma_j^y\Big)-\lambda\sum_{j=1}^N\sigma_j^z,
\end{eqnarray}
where $\sigma_j^{\alpha}$ is the Pauli matrix at site $j$ along $\alpha=x,y,z$ directions, $J$ is the ferromagnetic coupling in $x-y$ plane and $\lambda$ is the magnetic field along $z$ direction. Note that in addition to the Ising interaction $\sigma_i^x\sigma_j^x$ as written in Equation (10) of the main text, we consider a more general Hamiltonian including $\sigma_i^y\sigma_j^y$ interaction controlled by asymmetry $\gamma$. The model can be simplified by mapping to a large spin representation with definition $S_{\alpha}\equiv\sum_{j=1}^N\sigma_j^\alpha/2,\alpha=x,y,z$ and the large spin satisfies the commutation relations for spins, $[S_{\alpha},S_{\beta}]=i\epsilon_{\alpha\beta\gamma}S_{\gamma}$. In terms of large spin, the LMG Hamiltonian is
\begin{eqnarray}
\label{LMG}
H(\lambda)&=&-\frac{2J}{N}(S_x^2+\gamma S_y^2)-2\lambda S_z+\frac{(1+\gamma)J}{2},\\
&=&-\frac{J}{2N}(1-\gamma)(S_+^2+S_-^2)-\frac{J}{N}(1+\gamma)\Big(S^2-S_z^2-\frac{N}{2}\Big)-2\lambda S_z,
\end{eqnarray}
where the raising and lowering operator are defined as $S_{\pm}=S_x\pm iS_y$.  For $\gamma\neq1$, the LMG model presents a quantum phase transition at $\lambda_c=1$ from paramagnetic phase ($\lambda>1$) to ferromagnetic phase ($\lambda<1$). The critical exponents of the quantum phase transitions in the LMG model are $\nu=3/2,z=1/3$.

\subsection{I. Numerical Results of the Loschmidt echo in the LMG model}
\begin{figure}[t]
\centering
\includegraphics[width=\linewidth,angle=0]{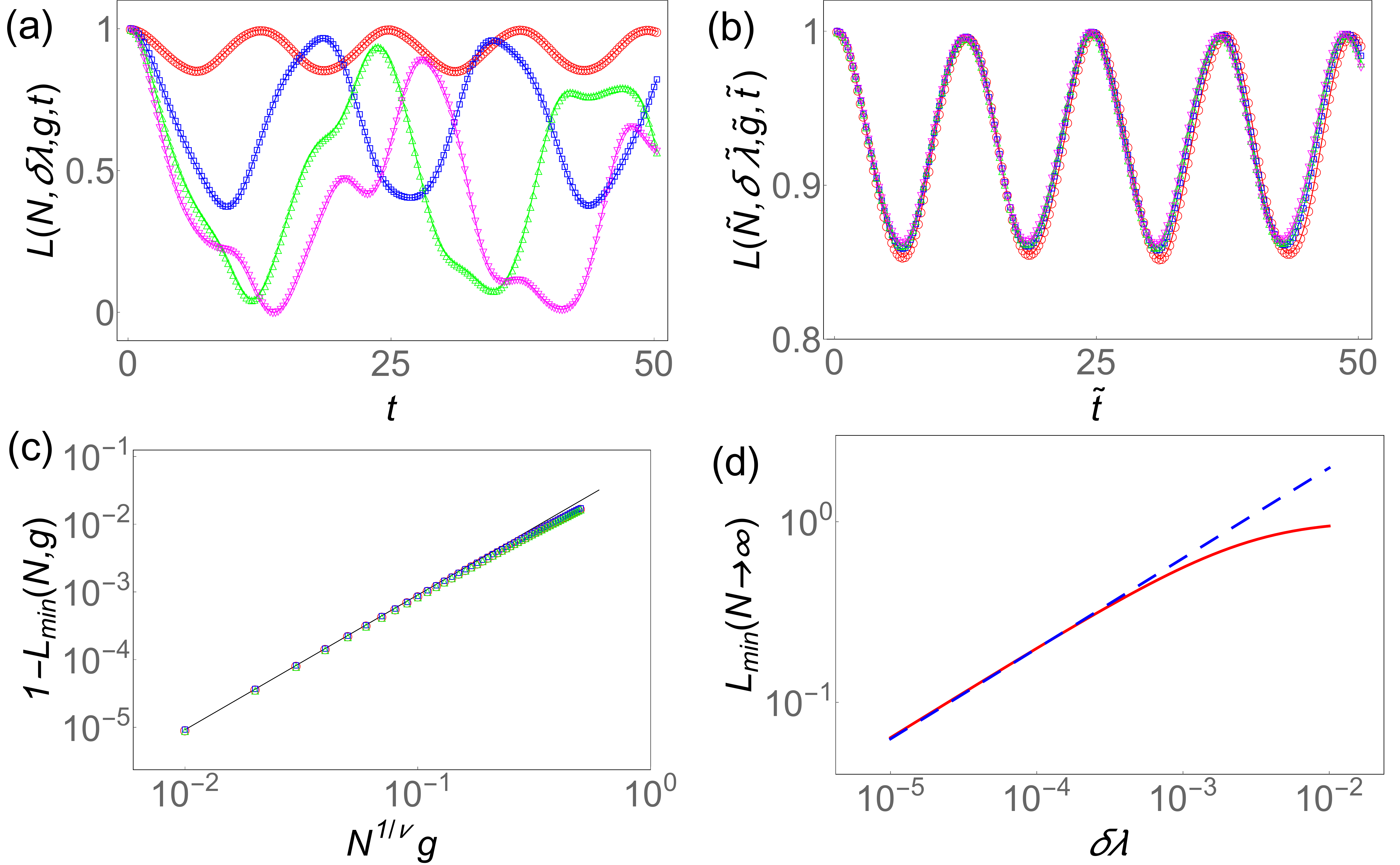}
\caption{Universal scaling of the Loschmidt echo in the LMG model. (a). The Loschmidt echo $L(N,\delta\lambda,g,t)$ with $\delta\lambda=1-\lambda_i=0.02$ and $g=0.01$ as a function of time for different system sizes, $N=500$ (the red circle), $N=1000$ (the blue square), $N=1500$ (the green upper triangle) and $N=2000$ (the magenta lower triangle). (b). Data collapse of the Loschmidt echo as a function of scale parameters $L(\tilde{N},\delta\tilde{\lambda},\tilde{g},\tilde{t})$ with $\tilde{N}=N/b$, $\delta\tilde{\lambda}=b^{1/\nu}\delta\lambda$, $g=b^{1/\nu}g$, $\tilde{t}=b^{-z}t$. We take $N=500,\delta\lambda=0.02,g=0.01$ and $\nu=3/2,z=1/3$ is exact critical exponents of the LMG model. $b=1$ (the red circle), $b=1/2$ (the blue square), $b=1/3$ (the green upper triangle) and $b=1/4$ (the magenta lower triangle). (c). Data collapse of $1-L_{\text{min}}(N,g)$ as a function of $N^{1/\nu}g$ with $\delta\lambda=0$. The red circle is for $N=500$ spins, the blue square is for $N=800$ spins and the green triangle for $N=1000$ spins. We can see that when $N^{1/\nu}g\ll1$, numerical fitting gives $1-L_{min}\propto (N^{1/\nu}g)^{\alpha}$ with $\alpha=1.99$, which confirms Equation (6) in the main text. (d). The scaling of the minimum of the Loschmidt echo in the thermodynamic limit as a function of $\delta\lambda$ (red solid line). Numerical fit of the data gives $L_{\text{min}}\propto\delta\lambda^{z\nu}$ with $z\nu=1/2$ (blue dashed line).}
\label{fig:S1}.
\end{figure}

As advertised in the main text, here we present numerical results for the LE dynamics of the LMG model in Figure~\ref{fig:S1}, which confirms all the predicted scaling laws. The Loschmidt echo decays and revivals in time for different system size $N$ is observed (Figure~\ref{fig:S1}-(a)). As the system sizes increases, the decays of Loschmidt echo are enhanced (Figure~\ref{fig:S1}-(a)). The scaling invariance of the LE presented in Equation (5) of the main text is confirmed in Figure~\ref{fig:S1}-(b) by showing the collapse of LE for a set of parameters related by the scaling transformation in Equation (3) of the main text. The first minimum of the Loschmidt echo at the quantum critical point $\delta\lambda=0$ for different values of $N$ and $g$ collapse to a universal function when rescaled as $N^{1/\nu}g$ (Figure~\ref{fig:S1}-(c)). In particular, in the limit of $N^{1/\nu}g\ll1$, we also see that $1-L_{\text{min}}$ follows a power-law $N^{2/\nu}$ as predicted in the main text(Figure~\ref{fig:S1}-(c)). When $\delta\lambda\ll g$, we show the scaling of the minimum of the Loschmidt echo in the thermodynamic limit as a function of $\delta\lambda$ (Figure~\ref{fig:S1}-(d)). In the limit of $\delta\lambda\rightarrow0$, we observe the minimum of Loschmidt echo decays as a power law $L_{\text{min}}\propto\delta\lambda^{z\nu}$ with $z\nu=1/2$ (Figure~\ref{fig:S1}-(d)).

\subsection{II. Diagonalization of the LMG model in the $N\rightarrow\infty$ Limit}
\noindent The LMG model is exactly solvable in the thermodynamic limit~\cite{Dusuel:2004eoa}. For completeness, we briefly review the diagonalization of Eq.~\eqref{LMG} through the Holstein-Primakoff (HP) transformation and the Bogoliubov transformation. The HP transformation maps the spin operators to boson operators,
\begin{eqnarray}
S_z&=&S-a^{\dagger}a,\\
S_+&=&\sqrt{2S-a^{\dagger}a}a, \label{hp1}\\
S_-&=&a^{\dagger}\sqrt{2S-a^{\dagger}a}. \label{hp2}
\end{eqnarray}
Here $a$ and $a^{\dagger}$ are the boson annihilation and creation operators, respectively and they satisfy the standard commutation relations for bosons, $[a,a^{\dagger}]=1,[a,a]=[a^{\dagger},a^{\dagger}]=0$.

In the thermodynamic limit, the magnetization of the ground state in both ordered and normal phase of the LMG can be obtained from the semi-classical solution, where the spin operator is represented by a vector in the three dimensional space. For $\lambda>1$, the spins are fully polarized along the $z$ direction. Therefore, one may choose the $z$ direction as the axis of quantization and thus the transformations defined in Equations \eqref{hp1} and \eqref{hp2} are valid. However, when $\lambda<1$, the spin aligns somewhere between $z$ axis and $x$ axis, whose angle is determined by $\alpha=\arccos(\lambda/J)$. Therefore, before applying the HP transformation, one has to bring the $z$ axis along the semi-classical magnetization direction by rotating the spin operators about the $y$ axis through
\begin{eqnarray}
\left(
  \begin{array}{c}
    S_x \\
    S_y \\
    S_z \\
  \end{array}
\right)=\left(
          \begin{array}{ccc}
            \cos\alpha & 0 & \sin\alpha \\
            0 & 1 & 0 \\
            -\sin\alpha & 0 & \cos\alpha \\
          \end{array}
        \right)\left(
  \begin{array}{c}
    \tilde{S}_x \\
    \tilde{S}_y \\
    \tilde{S}_z \\
  \end{array}
\right).
\end{eqnarray}
Here $\alpha=0$ for $\lambda>1$ and $\alpha= \cos^{-1} \lambda/J$ for $\lambda<1$. The Hamiltonian of the LMG model becomes,
\begin{eqnarray}
H(\lambda)&=&-\frac{2J}{N}\left[(\cos\alpha\tilde{S}_x+\sin\alpha\tilde{S}_z)^2+\gamma\tilde{S}_y^2\right]-2\lambda(-\sin\alpha\tilde{S}_x+\cos\alpha\tilde{S}_z)+\frac{(1+\gamma)J}{2}\label{Ham1}.
\end{eqnarray}
Considering the low energy excitations and setting $S = N/2$, HP transformation to the rotated spin operators becomes
\begin{eqnarray}
\tilde{S}_z&=&S-a^{\dagger}a=\frac{N}{2}-a^{\dagger}a,\\
\tilde{S}_+&=&\sqrt{2S-a^{\dagger}a}a=\sqrt{N}\sqrt{1-a^{\dagger}a/N}a,\\
\tilde{S}_-&=&a^{\dagger}\sqrt{2S-a^{\dagger}a}=\sqrt{N}a^{\dagger}\sqrt{1-a^{\dagger}a/N}.
\end{eqnarray}
In the large $N$ limit, we keep terms of order $N$, $\sqrt{N}$, $N^0$ and approximate $\sqrt{1-a^{\dagger}a/N}$ by 1 in the HP transformation. The Hamiltonian \eqref{Ham1}
is then transformed into
\begin{eqnarray}
H(\lambda)&=&-\frac{J}{2N}\left(\cos^2\alpha-\gamma\right)N\left(a^2+{a^{\dagger}}^2\right)-\frac{J}{N}\left(\cos^2\alpha+\gamma\right)\left[\frac{N}{2}\left(\frac{N}{2}+1\right)-\left(\frac{N}{2}-a^{\dagger}a\right)^2\right]\nonumber\\
&& -\frac{J\cos\alpha\sin\alpha}{N}\left[N\sqrt{N}(a+a^{\dagger})-\sqrt{N}(aa^{\dagger}a+a^{\dagger}aa+a^{\dagger}a^{\dagger}a+a^{\dagger}aa^{\dagger})\right]-\frac{2J\sin^2\alpha}{N}\left(\frac{N}{2}-a^{\dagger}a\right)^2\nonumber\\
&& +\lambda\sin\alpha\sqrt{N}(a+a^{\dagger})-2\lambda\cos\alpha\left(\frac{N}{2}-a^{\dagger}a\right)+\frac{J(1+\gamma)}{2},\\
&\approx& \left(-\frac{J\sin^2\alpha}{2}-\lambda\cos\alpha\right)N-\frac{J}{2}(\cos^2\alpha-1)+\frac{J(1+\gamma)}{2}+\sqrt{N}\sin\alpha(-J\cos\alpha+\lambda)(a+a^{\dagger})\nonumber\\
&& \left[-J(\cos^2\alpha+\gamma)+2J\sin^2\alpha+2\lambda\cos\alpha\right]a^{\dagger}a-\frac{J}{2}(\cos^2\alpha-\gamma)(a^2+{a^{\dagger}}^2).
\end{eqnarray}
For $\lambda>1$, $\cos\alpha=1$ and $\sin\alpha=0$, the LMG Hamiltonian is
\begin{eqnarray}
H(\lambda)&=&-\lambda N+\frac{J(1+\gamma)}{2}+\left[-J(1+\gamma)+2\lambda\right]a^{\dagger}a-\frac{J(1-\gamma)}{2}(a^2+{a^{\dagger}}^2).
\end{eqnarray}
For $\lambda<1$, $\cos\alpha=\lambda/J$ and $\sin\alpha=\sqrt{1-\lambda^2/J^2}$, the LMG Hamiltonian is
\begin{eqnarray}
H(\lambda)&=&-\frac{N}{2J}\left[J^2+\lambda^2\right]+\frac{1}{2J}\left(J^2-\lambda^2\right)+\frac{J(1+\gamma)}{2}+\frac{2J^2-\gamma J^2-\lambda^2}{J}a^{\dagger}a+\frac{J\gamma-\lambda^2}{2J}(a^2+{a^{\dagger}}^2).
\end{eqnarray}

\subsection{III. Analytic results of the Loschmidt echo in the $N\rightarrow\infty$ for $\lambda<1$}
\noindent For simplicity of notation, we take $J=1$ as the unit of the energy scale. For $\lambda<1$, the LMG Hamiltonian in the limit of $N\rightarrow\infty$ is
\begin{eqnarray}
H(\lambda)&=&-\frac{N}{2}\left[1+\lambda^2\right]+1+2(1-\gamma)a^{\dagger}a-\frac{(\lambda^2-\gamma)}{2}(a+a^{\dagger})^2.
\end{eqnarray}
The above Hamiltonian can be diagonalized by introducing the squeezing operator
\begin{eqnarray}
S(\xi)=\exp\left[\frac{1}{2}\left(\xi {a^{\dagger}}^2-\xi^*a^2\right)\right],
\end{eqnarray}
with $\xi(\lambda,\gamma)=-\frac{1}{4}\ln\frac{1-\lambda^2}{1-\gamma}$ being the squeezing parameter. The transformed Hamiltonian is
\begin{eqnarray}
S^{\dagger}(\xi)H(\lambda<1)S(\xi)=\epsilon_{ub}(\lambda)a^{\dagger}a,
\end{eqnarray}
where
\begin{eqnarray}
 \epsilon_{ub}=2(1-\gamma)e^{-2\xi(\lambda,\gamma)}=2\sqrt{(1-\gamma)(1-\lambda^2)}.
\end{eqnarray}
The ground state is then given by
\begin{eqnarray}
|\psi_0(\lambda)\rangle&=&S(\xi)|0\rangle=|\xi(\lambda,\gamma)\rangle.
\end{eqnarray}
Here $|0\rangle$ is the vacuum state for the bosons and $|\xi(\lambda,\gamma)\rangle$ is the squeezed state with parameter $\xi(\lambda,\gamma)$.

The Loschmidt echo for $\lambda_i<1$ and $\lambda_f<1$ ($\gamma$ is fixed) is defined by
\begin{eqnarray}
L(\lambda_i,\lambda_f,t)&=&\left|\langle\psi_0(\lambda_i)|e^{-itH(\lambda_f)}|\psi_0(\lambda_i)\rangle\right|^2.
\end{eqnarray}
Then we get
\begin{eqnarray}
L(\lambda_i,\lambda_f,t)&=&\left|\langle\xi(\lambda_i)|e^{-itH(\lambda_f)}|\xi(\lambda_i)\rangle\right|^2,\\
&=&\left|\langle\xi(\lambda_i)-\xi(\lambda_f)|(\xi(\lambda_i)-\xi(\lambda_f))e^{-2it\epsilon_{ub}(\lambda_f)}\rangle\right|^2.
\end{eqnarray}
The overlap between two squeezed states $|\xi_1=r_1e^{i\theta_1}\rangle$ and $|\xi_2=r_2e^{i\theta_2}\rangle$ is given by
\begin{eqnarray}
\langle\xi_2|\xi_1\rangle&=&\left[\cosh(r_1)\cosh(r_2)-e^{i(\theta_1-\theta_2)}\sinh(r_1)\sinh(r_2)\right]^{-1/2}.
\end{eqnarray}
Therefore the Loschmidt echo in the LMG model in the $N\rightarrow\infty$ for $\lambda_i<1,\lambda_f<1$ is
\begin{eqnarray}\label{LMGecho1}
L(\lambda_i,\lambda_f,t)&=&\left[\cosh^4(x)+\sinh^4(x)-2\sinh^2(x)\cosh^2(x)\cos(2t\epsilon_{ub}(\lambda_f))\right]^{-1/2},
\end{eqnarray}
where we define $x\equiv\xi(\lambda_i)-\xi(\lambda_f)$. Eq.\eqref{LMGecho1} means that the Loschmidt echo in the thermodynamic limit presents periodic oscillations in time with period
\begin{eqnarray}
T_u&=&\frac{\pi}{\epsilon_{ub}(\lambda_f)}=\frac{\pi}{2\sqrt{(1-\gamma)(1-\lambda_f^2)}}.
\end{eqnarray}
The maximum of Loschmidt echo in Eq.\eqref{LMGecho1} is always 1 and thus Loschmidt echo is fully recovered in one period. While the minimum of Loschmidt echo in Eq.\eqref{LMGecho1} occurs when $\cos(2t\epsilon_{ub}(\lambda_f))=-1$, i.e. the time $t=(n+1/2)T_u,n=0,1,2,\cdots$. The minimum of the Loschmidt echo is
\begin{eqnarray}
\label{lmg1}
L_{\min}(\lambda_i,\lambda_f)&=&\left(\cosh^2(x)+\sinh^2(x)\right)^{-1}=2\left(e^{2x}+e^{-2x}\right)^{-1}.
\end{eqnarray}
We note that we can write $x$ in terms of $\epsilon_{ub}$,
\begin{align}
x&=-\frac{1}{2}\log\epsilon_{ub}(\lambda_i)+\frac{1}{2}\log\epsilon_{ub}(\lambda_f),
\end{align}
from which we have
\begin{align}
e^{2x}=\frac{\epsilon_{ub}(\lambda_f)}{\epsilon_{ub}(\lambda_i)}.
\end{align}
For a coupling strength that is close enough to the critical point such that we have $\delta\lambda\ll g$, we have
\begin{align}
e^{2z_0}\propto|\delta\lambda|^{-z\nu}
\end{align}
where $\epsilon_{ub}(1-\delta\lambda)\propto |\delta\lambda|^{z\nu}$. Finally, we find the critical exponent of the minimum of the Loschmidt echo as
\begin{align}
\label{lmg2}
L_\textrm{min}(\delta\lambda)\propto\delta\lambda^{z\nu}.
\end{align}

\subsection{IV. Analytic results of the Loschmidt echo for the LMG model in the $N\rightarrow\infty$ for $\lambda>1$}
\noindent The Hamiltonian of the LMG model in the $N\rightarrow\infty$ for $\lambda>1$ is
\begin{eqnarray}
H&=&-\lambda N+1+2(\lambda-\gamma)a^{\dagger}a+\frac{(\gamma-1)}{2}(a+a^{\dagger})^2.
\end{eqnarray}
This Hamiltonian can also be diagonalized by introducing the squeezing operator
\begin{eqnarray}
S(\xi)=\exp\left[\frac{1}{2}\left(\xi {a^{\dagger}}^2-\xi^*a^2\right)\right],
\end{eqnarray}
where $\xi(\lambda,\gamma)=-\frac{1}{4}\ln\frac{\lambda-1}{\lambda-\gamma}$. The diagonalzied Hamiltonian is
\begin{eqnarray}
S^{\dagger}(\xi)H(\lambda>1)S(\xi)=\epsilon_{pp}(\lambda)a^{\dagger}a,
\end{eqnarray}
where $\epsilon_{pp}=2\sqrt{(\lambda-\gamma)(\lambda-1)}$. The ground state is then given by
\begin{eqnarray}
|\psi_0(\lambda)\rangle&=&S(\xi)|0\rangle=|\xi(\lambda,\gamma)\rangle.
\end{eqnarray}
Here $|0\rangle$ is the vacuum state for the bosons and $|\xi(\lambda,\gamma)\rangle$ is the squeezed state with parameter $\xi(\lambda,\gamma)$.

Similar to the case of $\lambda<1$, the analytic expression for the Loschmidt echo in the LMG model for $\lambda>1$ is
\begin{eqnarray}
L(\lambda_i,\lambda_f,t)&=&\left[\cosh^2(\xi(\lambda_0)-\xi(\lambda_1))-e^{2it\epsilon_{pp}(\lambda_1)}\sinh^2(\xi(\lambda_0)-\xi(\lambda_f))\right]^{-1}.
\end{eqnarray}
Then
\begin{eqnarray}\label{LMGecho2}
L(\lambda_i,\lambda_f,t)&=&\left[\cosh^4(x)+\sinh^4(x)-2\sinh^2(x)\cosh^2(x)\cos(2t\epsilon_{pp}(\lambda_f))\right]^{-1/2}.
\end{eqnarray}
Here we define $x\equiv\xi(\lambda_i)-\xi(\lambda_f)$. Eq.~\eqref{LMGecho2} means that the Loschmidt echo shows periodic oscillation in time with period
\begin{eqnarray}
T_p&=&\frac{\pi}{\epsilon_{pp}(\lambda_f)}=\frac{\pi}{2\sqrt{(\lambda_f-\gamma)(\lambda_f-1)}}.
\end{eqnarray}
The maximum of Loschmidt echo in Eq.~\eqref{LMGecho2} is always 1 and thus Loschmidt echo is fully recovered in one period. While the minimum of Loschmidt echo in Eq.~\eqref{LMGecho2} occurs when
$\cos(2t\epsilon_{pp}(\lambda_f))=-1$, i.e. $t=(n+1/2)T_p,n=0,1,2,\cdots$. The minimum of the Loschmidt echo is
\begin{eqnarray}
L_{\min}(\lambda_i,\lambda_f)&=&\left(\cosh^2(x)+\sinh^2(x)\right)^{-1}=2\left(e^{2x}+e^{-2x}\right)^{-1}.
\end{eqnarray}
Agian, by expressing $x$ in terms of $\epsilon_{pp}$, we have
\begin{align}
x&=-\frac{1}{2}\log\epsilon_{pp}(\lambda_i)+\frac{1}{2}\log\epsilon_{pp}(\lambda_f),
\end{align}
from which we have
\begin{align}
e^{2x}=\frac{\epsilon_{pp}(\lambda_f)}{\epsilon_{pp}(\lambda_i)}\propto|\delta\lambda|^{-z\nu}
\end{align}
for $\delta\lambda\ll g$ and where we have used $\epsilon_{pp}(1-\delta\lambda)\propto |\delta\lambda|^{z\nu}$. Finally, we find the critical exponent of the minimum of the Loschmidt echo as
\begin{align}
L_\textrm{min}(\delta\lambda)\propto\delta\lambda^{z\nu}.
\end{align}
Thus the minimum of Loschmidt echo at both sides ($\lambda>1$ and $\lambda<1$) has the same critical exponent $z\nu$.

\subsection{V. Ground state fidelity as a lower bound of Loschmidt echo}
Here we prove that the the Loschmidt echo is lower bounded by the ground state fidelity. The ground state fidelity between the initial and final state of the quench protocol reads
\begin{equation}
	F(\lambda_i, \lambda_f)=\braket{\psi_0(\lambda_i)|\psi_0(\lambda_f)}.
\end{equation}
Let us focus on only $\lambda<1$ (the same calculation is straightforward for $\lambda>1$). In the $N\rightarrow\infty$ limit, we have
\begin{align}
	F(\lambda_i, \lambda_f)&=\braket{\xi(\lambda_i)|\xi(\lambda_f)}=\cosh(x)^{-1/2}
\end{align}
By comparing this with Eq.~\eqref{lmg1}, we find that
\begin{align}
L_\textrm{min}(\lambda_i,\lambda_i+g)=(2 F(\lambda_i,\lambda_i+g)^{-4}-1)^{-1}.
\end{align}
Namely, we find here that a lower bound of the Loschmidt echo is determined by the ground state fidelity.

\subsection{VI. Finite-size scaling of $L_\textrm{min}$}
That the minimum of Loschmidt echo is bounded by the fidelity for which the scaling theory is well-known to be applicable~\cite{Rams:2011qu}, we use the scaling hypothesis to derive the finite-$N$ scaling exponent from Eq.~\eqref{lmg2}. Namely, let us now assume that there exists an analytical function with which $L_\textrm{min}$ is expressed as
\begin{equation}
L_\textrm{min}(\lambda_i,N)=|\delta\lambda|^{z\nu}\Phi(N^{1/\nu}|\delta\lambda|).
\end{equation}
with $\lim_{x\rightarrow\infty}\Phi(x)$ being a constant to recover Eq.~\eqref{lmg2}. For $N<\infty$, the non-analyticity at $\delta\lambda=0$ can be removed by requiring that
\begin{equation}
\lim_{x\rightarrow0}\Phi(x)\propto x^{-z\nu}.
\end{equation}
From this, we find
\begin{equation}
L_\textrm{min}(\lambda_i,N)\propto N^{-z}.
\end{equation}

\section{Section B: The Loschmidt echo in the quantum Rabi model}
\subsection{I. Critical scaling in the $\eta\rightarrow\infty$ limit}
We consider the normal phase of the QRM, $\lambda<1$, whose effective Hamiltonian is given by~\cite{Hwang:2015eq}
\begin{align}
H_{np}&=\omega_0a^\dagger a-\frac{\omega_0\lambda^2}{4}(a+a^\dagger)^2.
\end{align}
This can be diagonalized by using a squeezing operator $
\mathcal{S}[\xi]=\exp\left[\frac{1}{2}(\xi a^{\dagger2}-\xi^*a^2)\right]$
with $\xi=r_{np}(\lambda)$ where
\begin{align}
e^{r_{np}(\lambda)}=(1-\lambda^2)^{-\frac{1}{4}},\quad r_{np}(\lambda)=-\frac{1}{4}\ln(1-\lambda^2).
\end{align}
That is, we have
\begin{align}
\label{qrm1}
\mathcal{S}^\dagger[r] H_{np} \mathcal{S}[r]=\epsilon_{np}(\lambda)\adag a\end{align}
up to an additive constant and with
\begin{align}
\label{qrm2}
\epsilon_{np}(\lambda)=\omega_0e^{-2r}=\omega_0\sqrt{1-\lambda^2}.
\end{align}
The ground state is
\begin{align}
\ket{\phi_{0,np}(\lambda)}=\mathcal{S}[r_{np}]\ket{0}=\ket{r_{np}}.
\end{align}
The Loschmidt echo in the normal phase is
\begin{align}
L(\lambda_i,\lambda_f,t)=\left|\bra{\phi_{0,np}(\lambda_i)} \exp[-itH_{np}(\lambda_f)] \ket{\phi_{0,np}(\lambda_i)}\right|^2.
\end{align}
where$\lambda_f=\lambda_i+g$ and $\lambda_i=\lambda_c+\delta \lambda$.  First, using Eq.~\eqref{qrm1}, we find
\begin{align}
L(\lambda_i,\lambda_f,t)&=\left|\braket{r_{np}(\lambda_i)-r_{np}(\lambda_f)|(r_{np}(\lambda_i)-r_{np}(\lambda_f))e^{-2it\epsilon_{np}(\lambda_f)}}\right|^2\\
&=\left|\braket{z_0|z_0e^{-2it\epsilon_{np}(\lambda_f)}}\right|^2
\end{align}
where
\begin{align}
z_0\equiv r_{np}(\lambda_i)-r_{np}(\lambda_f).
\end{align}
Therefore, the analytical expression for the Loschmidt echo reads
\begin{align}
L(\lambda_i,\lambda_f,t)&=\left|\cosh^2(z_0)-e^{2it\epsilon_{np}(\lambda_f)}\sinh^2(z_0)\right|^{-1}\nonumber,\\
&=\left(\cosh^4(z_0)+\sinh^4(z_0)-2\cos(2t\epsilon_{np}(\lambda_f))\sinh^2(z_0)\cosh^2(z_0)\right)^{-\frac{1}{2}}.
\end{align}
Therefore, we conclude that the Loschmidt echo oscillates in time with a period $T=\pi/\epsilon_\textrm{np}(\lambda_f)$. The maximum value of the Loschmidt echo is always $1$, therefore, it is fully recovered in a full period. The minimum of the Loschmidt echo occurs when
\begin{align}
\cos(2t\epsilon_{np}(\lambda_f))=-1.
\end{align}
Namely, for $t=T/2+n\times T$ where $n$ is an integer. Let us now examine the scaling behavior of the minimum of the Loschmidt echo, which reads
\begin{align}
\label{qrm3}
L_\textrm{min}(\lambda_i,\lambda_f)&=\left(\cosh^4(z_0)+\sinh^4(z_0)+2\sinh^2(z_0)\cosh^2(z_0)\right)^{-\frac{1}{2}}\nonumber\\
&=\left(\cosh^2(z_0)+\sinh^2(z_0)\right)^{-1}\nonumber\\
&=2\left(e^{2z_0}+e^{-2z_0}\right)^{-1}.
\end{align}
Using Eq.~\eqref{qrm2}, we write $z_0$ in terms of $\epsilon_{np}$,
\begin{align}
z_0&=-\frac{1}{2}\log\epsilon_{np}(\lambda_i)+\frac{1}{2}\log\epsilon_{np}(\lambda_f),
\end{align}
from which we have
\begin{align}
e^{2z_0}=\frac{\epsilon_{np}(\lambda_f)}{\epsilon_{np}(\lambda_i)}
\end{align}
For a coupling strength that is close enough to the critical point such that we have $\delta\lambda\ll g$, we have
\begin{align}
e^{2z_0}\propto|\delta\lambda|^{-z\nu}
\end{align}
where $\epsilon_{np}(1-\delta\lambda)\propto |\delta\lambda|^{z\nu}$. Finally, we find the critical exponent of the minimum of the Loschmidt echo as
\begin{align}
\label{qrm4}
L_\textrm{min}(\delta\lambda)\propto\delta\lambda^{z\nu}.
\end{align}

\subsection{II. Ground state fidelity as a lower bound of Loschmidt echo}
Here we analytically show that the Loschmidt echo is lower bounded by the ground state fidelity. First, let us calculate the ground state fidelity between the initial and final state of the quench protocol,
\begin{equation}
	F(\lambda_i, \lambda_f)=\braket{\psi_0(\lambda_i)|\psi_0(\lambda_f)}.
\end{equation}
For the QRM model, in the $\eta\rightarrow\infty$ limit, we have
\begin{align}
	F(\lambda_i, g)&=\braket{r_{np}(\lambda_i)|r_{np}(\lambda_f)}=\cosh(r_{np}(\lambda_i)-r_{np}(\lambda_f))^{-1/2}=\cosh(z_0)^{-1/2}.
\end{align}
By comparing this with Eq.~\eqref{qrm3}, we find that
\begin{align}
L_\textrm{min}(\lambda_i,\lambda_i+g)=(2 F(\lambda_i,g)^{-4}-1)^{-1}.
\end{align}
Namely, we find here that a lower bound of the Loschmidt echo is determined by the ground state fidelity.

\subsection{III. Finite-$\eta$ scaling of $L_\textrm{min}$}
Following the same idea used for the LMG model, we use the scaling hypothesis to derive the finite-$\eta$ scaling exponent from Eq.~\eqref{qrm4}. Namely, let us now assume that there exists an analytical function with which $L_\textrm{min}$ is expressed as
\begin{equation}
L_\textrm{min}(\lambda_i,\eta)=|\delta\lambda|^{z\nu}\Phi(\eta^{1/\nu}|\delta\lambda|).
\end{equation}
with $\lim_{x\rightarrow\infty}\Phi(x)$ being a constant to recover Eq.~\eqref{qrm4}. For $\eta<\infty$, the non-analyticity at $\delta\lambda=0$ can be removed by requiring that
\begin{equation}
\lim_{x\rightarrow0}\Phi(x)\propto x^{-z\nu}.
\end{equation}
From this, we find
\begin{equation}
L_\textrm{min}(\lambda_i,\eta)\propto \eta^{-z}.
\end{equation}

\subsection{IV. Analytical approach for the scaling invariance of the Loschmidt echo}
For a finite but large $\eta$, the low-energy effective Hamiltonian of the QRM becomes~\cite{Hwang:2015eq}
\begin{align}
H_{np}(\lambda,\eta)&=\omega_0a^\dagger a-\frac{\omega_0\lambda^2}{4}(a+a^\dagger)^2+\frac{\omega_0\lambda^4}{16}\frac{1}{\eta}(a+a^\dagger)^4.
\end{align}
with the quartic correction term. The Loschmidt echo is
\begin{align}
L(\lambda_i, \lambda_f,\eta,t)=\left|\bra{\phi_{0,np}(\lambda_i,\eta)} \exp[-itH_{np}(\lambda_f,\eta)] \ket{\phi_{0,np}(\lambda_i,\eta)}\right|^2.
\end{align}
As shown in Ref.~\cite{Hwang:2015eq}, the ground state $\ket{\phi_{0,np}(\lambda_i,\eta)}$ is very well-approximated by a squeezed state, whose squeezing parameter $r(\lambda_i,\eta)$ is determined by a following equation
\begin{align}
\frac{3}{2}\frac{\lambda_i^4}{\eta}\xi^3+(1-\lambda_i)(1+\lambda_i)\xi^2-1=0,
\end{align}
where $\xi=e^{2r(\lambda_i,\eta)}$. For $\delta\lambda\ll1$, the above  equation becomes invariant under a following transformation
\begin{align}
\label{qrm6}
\eta\rightarrow\eta/b,\quad, \delta\lambda\rightarrow b^{2/3},\quad\xi\rightarrow\xi/b^{1/3}.
\end{align}
On the other hand, on the order of $\mathcal{O}[\eta^{-1}]$, the Loschmidt echo for $\delta\lambda=0$ for $\eta\gg1$ and $g \ll 1$ has the identical structure as the $\eta\rightarrow\infty$ case, but with a modified squeezing parameter $r(\lambda_i,\eta)$. That is,
\begin{align}
L(\eta,g,t)=\left|\braket{z|ze^{-2it\epsilon_{np}(\eta,g)}}\right|^2
\end{align}
where
\begin{align}
z=r(\lambda_c,\eta)-r(\lambda_c-g,\eta).
\end{align}
Using the equilibrium scaling properties of $\epsilon_\textrm{np}(\eta,g)$~\cite{Hwang:2015eq} where $F_\epsilon$ is a scaling function, we write
\begin{align}
L(\eta,g,t)=\left|\braket{z|ze^{-2it\eta^{-1/3}F_\epsilon(g \eta^{2/3})}}\right|^2
\end{align}
Now, we have all the ingredients to show the scaling invariance of the LE. First, the oscillating term is invariant under
\begin{align}
\label{qrm7}
\eta\rightarrow\eta/b,\quad\delta\lambda\rightarrow \delta \lambda b^{2/3}, \quad t\rightarrow t/b^{1/3}
\end{align}
Furthermore, the parameter $z$ becomes invariant under
\begin{align}
\label{qrm7}
\eta\rightarrow\eta/b,\quad\delta\lambda\rightarrow \delta \lambda b^{2/3}, \quad g\rightarrow gb^{2/3}.
\end{align}
This is because, upon the above transformation, we have $r(\lambda_c,\eta)\rightarrow r(\lambda_c,\eta)-\frac{1}{6}\log b $ and $r(\lambda_c-g,\eta)\rightarrow r(\lambda_c-g,\eta)-\frac{1}{6}\log b$ and therefore the constant shift from the scaling transformation cancels out for $z$. By combining Eq.~\eqref{qrm6} and \eqref{qrm7}, we therefore prove that
\begin{align}
L(\eta,\delta\lambda,g,t)=L(Nb^{-1},\delta\lambda b^{2/3},g b^{2/3},t b^{-1/3}),
\end{align}
which agrees with the result in the main text with $\nu=3/2$ and $z=1/3$.

\section{Section C: Loschmidt echo in the transverse field Ising Chain}
\noindent The analytical expression for the LE of TFIC is given in Ref.~\cite{Quan:2006fda}. For a completeness, here we provide a derivation of the LE, which is used for a numerical evaluation of LE presented in the main text. The Hamiltonian of the transverse field Ising chain is
\begin{eqnarray}\label{Hising}
H&=&-J\sum_{i=1}^N[\sigma_i^x\sigma_{i+1}^x+\lambda\sigma_j^z].
\end{eqnarray}
For simplicity, we set $J=1$. We impose periodic boundary condition for the Pauli spins, namely $\sigma_{N+1}^{\alpha}=\sigma_1^{\alpha}$ with $\alpha=x,y,z$. Note that Hamiltonian of TFIC has a parity symmetry $P=\prod_j\sigma_j^z$, which can take two values $+1$ or $-1$, thus the Hamiltonian of the TFIC is block diagonalized in the subspace of $P$.

Performing the Jordan-Wigner transformations~\cite{Suzuki:2012qu}
\begin{eqnarray}
\sigma_j^z&=&1-2c_j^{\dagger}c_j \\
\sigma_j^{\dagger}&=&\prod_{i<j}(1-2c_i^{\dagger}c_i)c_j=c_j\prod_{i<j}\sigma_i^z\\
\sigma_j^{-}&=&\prod_{i<j}(1-2c_i^{\dagger}c_i)c_j^{\dagger}=c_j^{\dagger}\prod_{i<j}\sigma_i^z
\end{eqnarray}
where $c_i$ and $c_i^{\dagger}$ are spinless fermion operators which obey
anti-commutation relations $\{c_i,c_j^{\dagger}\}=\delta_{ij}$ and
$\{c_i,c_j\}=\{c_i^{\dagger},c_j^{\dagger}\}=0$, then Hamiltonian in Eq.~\eqref{Hising} becomes a fermion model
\begin{eqnarray}\label{Hfree}
H&=&-\sum_{j=1}^{N}\Big[(c_j^{\dagger}c_{j+1}+c_{j+1}^{\dagger}c_j)+(c_{j}^{\dagger}c_{j+1}^{\dagger}
+c_{j+1}c_j)\Big]-\lambda\sum_{j=1}^{N}(1-2c_j^{\dagger}c_j).
\end{eqnarray}
Note that in the invariant subspace of $P=1$, the fermion model Eq.~\eqref{Hfree} must take anti-periodic boundary condition $c_{N+1}=-c_1$. While in the invariant subspace of $P=-1$, the fermion model Eq.~\eqref{Hfree} must take periodic boundary condition $c_{N+1}=c_1$.

The transformed fermion model Eq.~\eqref{Hfree} is quadratic form, which can be diagonalized in the
momentum space. Performing Fourier transformation with definition
\begin{eqnarray}
c_j=\frac{1}{\sqrt{N}}\sum_{k}c_ke^{ikj},
\end{eqnarray}
the Hamiltonian in Equation \eqref{Hfree} becomes
\begin{eqnarray}
H&=&\sum_kH_k=\sum_{k>0}\left[2(\lambda-\cos k)(c_k^{\dagger}c_k+c_{-k}^{\dagger}c_{-k}-1)-2i\sin k(c_k^{\dagger}c_{-k}^{\dagger}-c_{-k}c_{k})\right].
\end{eqnarray}
The momentum $k$ is fixed by the boundary condition of the fermion model. In the subspace, $P=1$, anti-periodic boundary is imposed for the fermions, thus $k_j=\frac{(2j+1)\pi}{N}$, where $j=-\frac{N}{2},-\frac{N}{2}+1,\cdots,\frac{N}{2}-1$ if $N$
is even and $j=-\frac{N-1}{2},\cdots,\frac{N-1}{2}$ if $N$ is odd. While in the subspace of $P=-1$, the fermion model should take periodic boundary condition, the momentum is $k_j=\frac{2\pi j}{N}$, where $j=-\frac{N}{2},-\frac{N}{2}+1,\cdots,\frac{N}{2}-1$ if $N$
is even and $j=-\frac{N-1}{2},\cdots,\frac{N-1}{2}$ if $N$ is odd.

Different modes of $H_k$ commute and then we only need to diagonalize each mode separately. For a specific mode $k$, the Hilbert space of $H_k$ is four dimensional with the basis $|n_{k},n_{-k}\rangle$ are
$|0,0\rangle,|1,0\rangle,|0,1\rangle,|1,1\rangle$. Two basis states $|1,0\rangle$ and $|0,1\rangle$ are already eigenstates of $H_k$ with eigenvalue $0$. The remaining two states $|0,0\rangle,|1,1\rangle$ form an invariant subspace of $H_k$. We thus can define a pseudo Pauli-spin $\tau$ with the basis are $|\downarrow\rangle=|0,0\rangle$ and $|\uparrow\rangle=|1,1\rangle$.
Thus we have,
\begin{eqnarray}
H_k|\uparrow\rangle&=&2(\lambda-\cos k)|\uparrow\rangle+2i\sin k|\downarrow\rangle,\\
H_k|\downarrow\rangle&=&-2(\lambda-\cos k)|\downarrow\rangle-2i\sin k|\uparrow\rangle.
\end{eqnarray}
For simplicity of notation, we define $A_k=2(\lambda-J\cos k)$ and $B_k=2J\sin k$. In the invariant subspace, the Hamiltonian $H_k$ can be written as
\begin{eqnarray}
H_k=
\begin{pmatrix}
A_k & -iB_k\\
iB_k & -A_k
\end{pmatrix}
\end{eqnarray}
In terms of pseudo spin, we have
\begin{eqnarray}
H_k&=&A_k\tau_z+B_k\tau_y=\sqrt{A_k^2+B_k^2}\left[\frac{A_k}{\sqrt{A_k^2+B_k^2}}\tau_z+\frac{B_k}{\sqrt{A_k^2+B_k^2}}\tau_y\right]=\epsilon_k\left[e^{i\theta_k\tau_x/2}\tau_ze^{-i\theta_k\tau_x/2}\right].
\end{eqnarray}
Here we define $\epsilon_k=\sqrt{A_k^2+B_k^2}=2\sqrt{\lambda^2-2\lambda\cos k+1}$ and $\cos\theta_k=\frac{A_k}{\sqrt{A_k^2+B_k^2}}$ and $\sin\theta_k=\frac{B_k}{\sqrt{A_k^2+B_k^2}}$. Thus the eigenenergy of mode $k$ are $\pm2\sqrt{\lambda^2-2\lambda\cos k+1}$. In the matrix form, the Hamiltonian of mode $k$ is written as
\begin{eqnarray}
H_k&=&\begin{pmatrix}
\cos\frac{\theta_k}{2} & i\sin\frac{\theta_k}{2}\\
i\sin\frac{\theta_k}{2} & \cos\frac{\theta_k}{2}
\end{pmatrix}\begin{pmatrix}
\epsilon_k & 0\\
0 & -\epsilon_k
\end{pmatrix}\begin{pmatrix}
\cos\frac{\theta_k}{2} & -i\sin\frac{\theta_k}{2}\\
-i\sin\frac{\theta_k}{2} & \cos\frac{\theta_k}{2}
\end{pmatrix}
\end{eqnarray}
Thus the Loschmidt echo at zero temperature is
\begin{eqnarray}
L(t)&=&\left|\langle\Psi_0(\lambda_i)|e^{-itH(\lambda_f)}|\Psi_0(\lambda_i)\rangle\right|^2,\\
&=&\prod_{k>0}\left|\left(-i\sin\frac{\theta_k^i}{2},\cos\frac{\theta_k^i}{2}\right)
\begin{pmatrix}
\cos(\epsilon_k^ft)-i\sin(\epsilon_k^ft)\cos\theta_k^f & -\sin(\epsilon_k^ft)\sin\theta_k^f\\
\sin(\epsilon_k^ft)\sin\theta_k^f & \cos(\epsilon_k^ft)+i\sin(\epsilon_k^ft)\cos\theta_k^f
\end{pmatrix}\begin{pmatrix}
i\sin\frac{\theta_k^i}{2}\\
\cos \frac{\theta_k^i}{2}
\end{pmatrix}\right|^2,\\
&=&\prod_{k>0}\left|\left(-i\sin\frac{\theta_k^i}{2},\cos\frac{\theta_k^i}{2}\right)\begin{pmatrix}
i\cos(\epsilon_k^ft)\sin\frac{\theta_k^i}{2}-\sin(\epsilon_k^ft)\sin\left(\theta_k^f-\frac{\theta_k^i}{2}\right)\\
\cos(\epsilon_k^ft)\cos\frac{\theta_k^i}{2}+i\sin(\epsilon_k^ft)\cos\left(\theta_k^f-\frac{\theta_k^i}{2}\right)
\end{pmatrix}\right|^2,\\
&=&\prod_{k>0}\left|\cos(\epsilon_k^ft)+i\sin(\epsilon_k^ft)\cos(\theta_k^f-\theta_k^i)\right|^2.
\end{eqnarray}
Here we use several shorthand notation, $\epsilon_k^f\equiv\epsilon_k(\lambda_f)$, $\epsilon_k^i\equiv\epsilon_k(\lambda_i)$ and  $\theta_k^f\equiv\theta_k(\lambda_f)$ and $\theta_k^i\equiv\theta_k(\lambda_i)$.

\begin{figure}[t]
\centering
\includegraphics[width=\linewidth,angle=0]{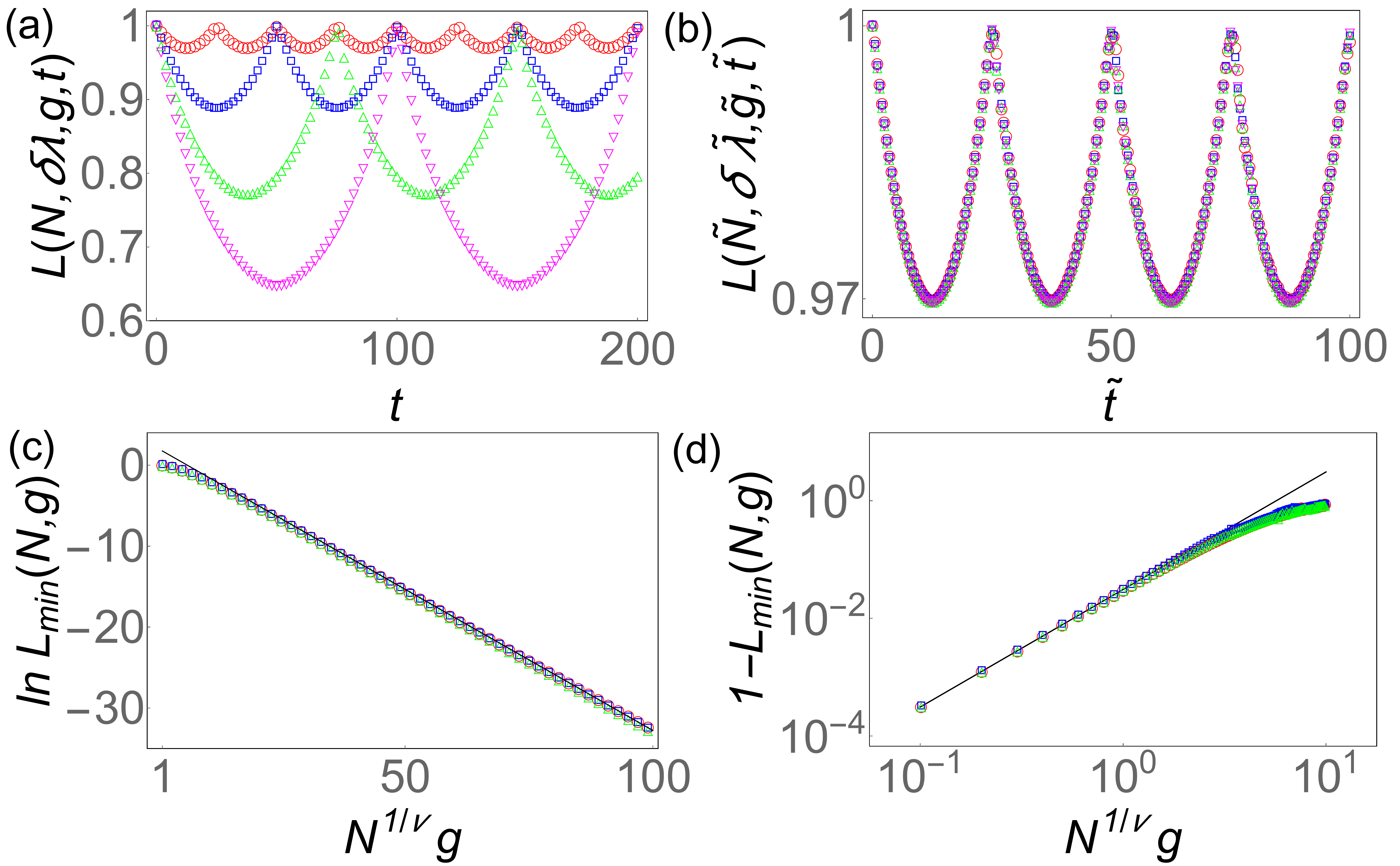}
\caption{Periodic oscillation of the Loschmidt echo (LE) in the TFIM. (a). The LE $L(N,\delta\lambda,g,t)$ with $\delta\lambda=0$ and $g=0.01$ as a function of time for different system sizes, $N=100$, $200$, $300$ and $400$ (from top to bottom). (b) Scaling invariance of LE given in Eq.~\eqref{eq04}. We take $N=100,\delta\lambda=0,g=0.01$ and $\nu=z=1$. All the data for $b=1$, $1/2$, $1/3$, and $1/4$ collapses onto a single curve. (c,d) The data collapse of the LE minimum as a function of $N^{1/\nu}g$ with $\delta\lambda=0$ for different values of $N$ and $g$ satisfying (c) $N^{1/\nu}g\gg1$ and (d) $N^{1/\nu}g\ll1$.}
\label{fig:S2}.
\end{figure}

Finally, we note that the decay and revival of the Loschmidt echo shown in Fig. 1 of the main text is for a quench protocol where both the initial and final parameters are different from the critical point. This leads to the irregularity in the oscillation (not perfectly periodic). When either the initial or final parameter of the quench protocol is chosen to be at the critical point, the Loschmidt echo exhibits a periodic oscillation as shown in Fig.~\ref{fig:S2}, which is similar to numerical results shown in Ref.~\cite{Quan:2006fda}. We have shown the irregular dynamics in the main text to show the robustness of the scaling invariance of the LE.

\section{Section D: The Su-Schrieffer-Heeger (SSH) model}
\noindent The Hamiltonian of the SSH model~\cite{Su:1979so},
\begin{eqnarray}\label{Hssh}
H&=&\sum_{j=1}^N\left[J_1a_j^{\dagger}b_j+J_2b_j^{\dagger}a_{j+1}+H.c.\right],
\end{eqnarray}
where $a_j$ and $b_j$ are the annihilation operators in two sublattices at site $j$, $J_1$ and $J_2$
are hopping amplitudes and we define a dimensionless control parameter $\lambda=J_2/J_1$. It is well know that $\lambda_c=1$ is the critical point of a topological phase transitions.

Performing a Fourier transformation with the definition, $a_j=\frac{1}{\sqrt{N}}\sum_ka_ke^{ikj},b_j=\frac{1}{\sqrt{N}}\sum_kb_ke^{ikj}$, the Hamiltonian in Equation \eqref{Hssh} becomes
\begin{eqnarray}
H&=&\sum_k\left[(J_1+J_2e^{-ik})a_k^{\dagger}b_k+(J_1+J_2e^{ik})b_k^{\dagger}a_k\right].
\end{eqnarray}
Introducing the spinor $\Gamma^{\dagger}=(a_k^{\dagger},b_k^{\dagger})$, the Hamiltonian can be written as
\begin{eqnarray}
H&=&\sum_{k}\Gamma^{\dagger}H_k\Gamma.
\end{eqnarray}
Here $H_k$ is given by
\begin{eqnarray}
H_k=\left(
  \begin{array}{cccc}
    0 & (J_1+J_2e^{-ik}) \\
    (J_1+J_2e^{ik}) & 0
  \end{array}
\right)
\end{eqnarray}
For periodic boundary conditions, $k=2\pi m/N,m=-N/2,\cdots,0,1,2,\cdots,N/2$. We define a pseudo Pauli-spin with the basis states $|\uparrow\rangle=|1_a,0_b\rangle,|\downarrow\rangle=|0_a,1_b\rangle)$. In terms of pseudo spins, the $H_k$ can be written as
\begin{eqnarray}
H_k&=&(J_1+J_2\cos k)\sigma_x+(J_2\sin k)\sigma_y.
\end{eqnarray}
Thus the eigenvalues of $H_k$ are respectively
\begin{eqnarray}
E_{\pm}(k)&=&\pm\sqrt{J_1^2+J_2^2+2J_1J_2\cos k}.
\end{eqnarray}
The corresponding eigenstates are respectively
\begin{eqnarray}
|\Psi_+(k)\rangle&=&\frac{1}{\sqrt{2}}\left(
                   \begin{array}{c}
                     e^{i\theta_k} \\
                     1 \\
                   \end{array}
                 \right)
\end{eqnarray}
\begin{eqnarray}
|\Psi_-(k)\rangle&=&\frac{1}{\sqrt{2}}\left(
                   \begin{array}{c}
                     -e^{i\theta_k} \\
                     1 \\
                   \end{array}
                 \right)
\end{eqnarray}
Here the $\theta_k$ is defined by the relation $(J_1+J_2e^{-ik})=E_+(k)e^{i\theta_k}$.

At zero temperature, the Loschmidt echo in the SSH model is
\begin{eqnarray}
L(t)&=&|\langle\psi_0(\lambda_i)|e^{-itH(\lambda_f)}|\psi_0(\lambda_i)\rangle|^2.
\end{eqnarray}
In the SSH, we define $\lambda=J_2/J_1$ and take $J_1=1$. Thus we have
\begin{eqnarray}
L(\lambda_i,\lambda_f,t)&=&\prod_{k}\left|\cos[tE_+(\lambda_f,k)]+i\sin[tE_+(\lambda_f,k)]\frac{[(1+\lambda_i\cos k)(1+\lambda_f\cos k)+\lambda_i\lambda_f\sin^2 k]}{E_+(\lambda_f,k)E_+(\lambda_i,k)}\right|^2.
\end{eqnarray}

\end{document}